# Power Balance in the ITER Plasma and Divertor

D. E. Post[1], B. Braams[2], J. Mandrekas[3], W. Stacey[3] and N. Putvinskaya[4]

[1]ITER Joint Central Team, ITER San Diego Co-Center, San Diego, California
[2]Courant Institute, New York University, New York, New York
[3]Georgia Institute of Technology, Atlanta, Georgia
[4]SAIC, ITER San Diego Joint Work Site

Abstract

It is planned to use atomic processes to spread out most of the heating power over the first wall and side walls to reduce the heat loads on the plasma facing components in ITER to ~ 50 MW. Calculations indicate that there will be 100 MW in bremsstrahlung radiation from the plasma center, 50 MW of radiation from the plasma edge inside the separatrix and 100 MW of radiation from the scrape-off layer and divertor plasma, leaving 50 MW of power to be deposited on the divertor plates. The radiation losses are enhanced by the injection of impurities such as Neon or Argon at acceptably low levels (~0.1 % Argon, etc.)

1. Introduction

Power exhaust will be a key issue for the next generation of fusion experiments such as ITER[1]. The alpha heating power in ITER will be about 300 MW, with possibly another 100 MW of auxiliary power for driven operation. This gives a peak heat load of about 30 MW/m$^2$ or larger. To maintain the surface temperature of the divertor plates at an acceptable level requires that the peak heat loads be reduced to ~ 5 MW/m$^2$.which requires that the total heat loads on the divertor plates be 50 MW or less. This can be accomplished by using impurity radiation, enhanced by the injection of impurities such as Ne or Ar, to spread out the plasma energy on the first wall and divertor chamber walls. An analysis of the various channels for energy losses from the plasma for the plasma core, the plasma edge inside the separatrix (mantle), the scrape-off layer and divertor plasma, and the divertor plates indicates that it should be possible to radiate about 250 MW of power to the walls before the energy reaches the plate (Figure 1 and Table 1).

Table 1. Power flow in ITER

| | |
|---|---|
| Bremsstrahlung | 100 MW |
| Edge impurity radiation (inside separatrix) | 50 MW |
| SOL/divertor impurity radiation (outside separatrix) + H and He radiation and charge exchange losses | 100 (up to 250) MW |
| Power on target plates | 50 MW |
| total heating power | 300 (up to 400) MW |

2. Detailed Power Losses in ITER

Bremsstrahlung losses will be larger in ITER than in present experiments due to the higher electron density, temperature and energy confinement time. The ratio of the Bremsstrahlung emission $P_{Brem}$ to the alpha heating $P_\alpha$ is independent of size and density (equation 1), where $C_B$ is $2\times 10^{-40}$ Zeff MW/(m$^6$eV$^{0.5}$), $n_e$ and $T_e$ are the volume averaged density and temperature, $f_{DT}=n_{DT}/n_e$, $E_\alpha$ = 3.5 MeV, and $\langle\sigma v\rangle_{DT}$ is the fusion reaction rate. For ITER, ($\langle T\rangle \sim 10$ keV, $f_\alpha \sim 0.1$, $f_{Be}\sim 0.01$, $f_{Ne}\sim 0.003$ and $P_{Brem}/P_\alpha \approx 0.33$ (Fig.2). $P_{Brem}/P_\alpha$ decreases with $T_e$ as $T_e^{-1.5}$, and increases with impurity level (eq.2) where $f_Z=n_{impurity}/n_e$.

$$\frac{P_{Brem}}{P_\alpha} = \frac{C_B\, n_e^2 Z_{eff} T^{1/2}}{n_e^2/4\ f_{DT}^2 \langle\sigma v\rangle_{DT} E_\alpha} = \frac{4 C_B Z_{eff} T^{1/2}}{f_{DT}^2 \langle\sigma v\rangle_{DT} E_\alpha} \qquad (1)$$

$$\frac{P_{Brem}}{P_\alpha} \propto \frac{Z_{eff}}{f_{DT}^2} \approx \frac{1+2f_\alpha+\sum Z(Z-1)f_Z}{1-2f_\alpha-\sum Z f_Z} \qquad (2)$$

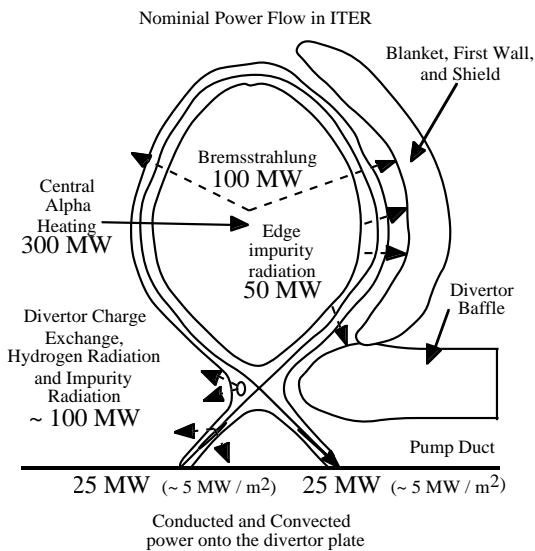

Figure 1. Power flow in ITER.

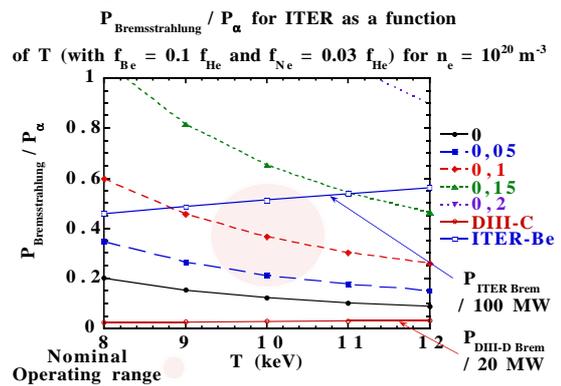

Figure 2. Ratio of Bremsstrahlung radiation to alpha heating as a temperature for $n_{He}/n_e$ = 0, 0.05, 0.1, 0.15 and 0.2. $P_{Brem}/P_{heat}$ is also plotted for $P_{heat}$ = 20 MW for DIII-D and 100 MW for ITER.

There will be impurity radiation from the main plasma edge (mantle) just inside the separatrix to the first wall due to line emission from intrinsic impurities and injected impurities such as Neon or Argon. This radiation can be calculated from the radial heat conduction equation and the radiation emission rates for candidate injected impurities (equation 3). If one assumes that the density and the electron transport coefficients are roughly constant, as is approximately true for H-mode plasmas [2], the radiation losses are proportional to an integral of the radiation emission rate[3,4](Figure 3). The effect of heat conduction on the radiation losses and the different scaling with density and impurity fraction is significant. With conduction, $Q_\perp \approx n_e \sqrt{f_z}$, which is much weaker than the volume loss rate $\approx n_e^2 f_z$.

$$Q_\perp = \kappa \frac{\partial T}{\partial r} \; ; \; \frac{\partial Q_\perp}{\partial r} = -n_e n_z L_z(T_e) \Rightarrow Q_\perp \frac{\partial Q_\perp}{\partial r} = -n_e^2 f_z L_z(T_e) \kappa \frac{\partial T}{\partial r}$$

$$\Rightarrow Q_{\perp center}^2 - Q_{\perp separatrix}^2 \approx 2 \int_{T_{separatrix}}^{T_{center}} n_e^2 f_z L_z(T_e) \kappa \, dT_e \approx 2 n_e^2 f_z \kappa \int_{T_{separatrix}}^{T_{center}} L_z(T_e) \, dT_e \quad (3)$$

For the range of expected conditions for ITER for Zeff ≈ 1.6 with $f_{He}$ = 0.1, $f_{Be}$ = 0.01, $8 \times 10^{19}$ m$^{-3}$ ≤ $n_{edge}$ ≤ $10^{20}$ m$^{-3}$, $0.125 \times 10^{20}$ m$^{-1}$ s$^{-1}$ ≤ κ ≤ $2 \times 10^{20}$ m$^{-1}$ s$^{-1}$, Neon, Argon and Krypton can provide radiation losses of about 50 MW from the plasma edge ($Q_\perp$ ~ 0.05 MW/m$^2$, Table 2). These analytic estimates are consistent with detailed calculations using the WHIST 1 1/2 tokamak transport code with a multi-species impurity transport package with the constraint that $\Delta P_\alpha / P_\alpha$ ≤ 5% and $\tau_E \approx 2 \times \tau_E^{ITERP-89}$.

Table 2 Radial Impurity Radiation losses for Be, C, Ne, Ar and Kr for ITER conditions

|  | Be | C | Ne | Ar | Kr |
|---|---|---|---|---|---|
| Assumed fraction ($Z_{eff}$ ≤ 1.6) | 0.01 | 0.01 | 0.0033 | 0.001 | .00025 |
| $Q_\perp / (n_e (\kappa_\perp f_z)^{0.5})$ MW m* | 0.1 | 0.4 | 1.0 | 3.0 | 10 |
| $Q_\perp$ (MW/m$^2$) for $5 \times 10^{19}$ ≤ $n_e$ < $10^{20}$ and 0.125 ≤ κ ≤ 2 | 0.002 to 0.014 | 0.007 to 0.06 | 0.01 to 0.08 | 0.02 to 0.13 | 0.03 to 0.22 |

*$\kappa_\perp$ in $10^{20}$ m$^{-1}$ s$^{-1}$, $n_e$ in $10^{20}$ m$^{-3}$

A similar model can be used to estimate the radiation losses from the scrape-off layer and divertor plasma [4,5]( equation 4). The parallel heat conduction is classical and the pressure is constant along the field lines. The radiation efficiency can be computed as a function of the upstream temperature(Figure 5). The upstream temperature can be calculated by integrating the heat flux equation from the midplane to the radiation region. The impurity radiation losses can be computed for candidate ITER upstream densities (Table 3). For heating powers of 100 to 200 MW, $Q_\parallel \approx$ 0.3 to 0.6 GW/m$^2$ for each divertor leg. Neon and higher

Z impurities can radiate 100 MW or more from the divertor and SOL plasma. These losses will be supplemented by H, He and Be radiation and charge exchange losses. These results are consistent with detailed 2 D B2 code divertor modelling calculations using a multi-species transport model for Neon or Argon impurities (Table 4).

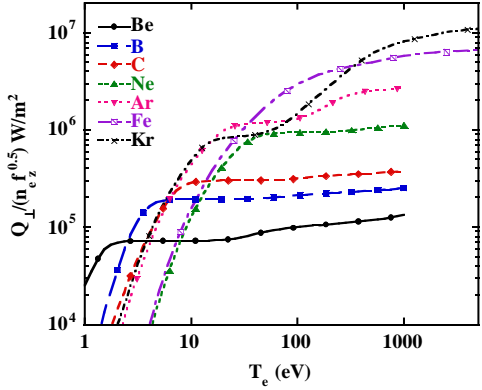

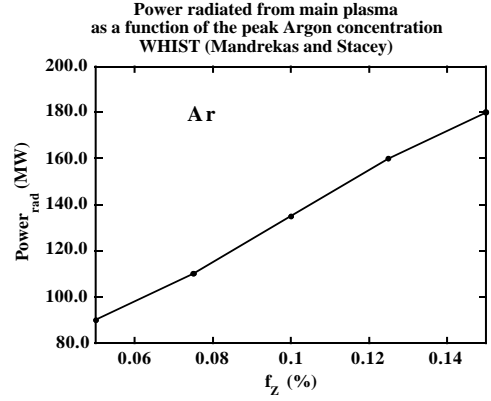

Figure 3. The normalized radiation losses $Q_\perp / n_e f_z^{0.5} \kappa_z^{0.5} = \sqrt{2 \int_0^T L_z(T_e)\, dT_e}$ vs. $T_e$.

Figure 4. Power radiated from the main plasma as a function of the peak Argon concentration from a Whist calculation.

$$\frac{\partial Q_\parallel}{\partial x} = -n_e n_z L_Z(T_e) \ ; \ Q_\parallel = -\kappa_o T_e^{2.5}\frac{\partial T_e}{\partial x} \ ; \ p_e = n_e T_e \ \Rightarrow$$

$$\frac{1}{2}\frac{\partial Q_\parallel^2}{\partial x} \approx \frac{p_e^2}{T_e^2}\kappa_o f_z L_Z T_e^{2.5}\frac{\partial T_e}{\partial x} \approx p_e^2 \kappa_o f_z L_Z T_e^{0.5}\frac{\partial T_e}{\partial x} \ \Rightarrow \ \frac{1}{2} dQ_\parallel^2 \approx p_{es}^2 \kappa_o f_z L_Z T_e^{0.5} dT_e \quad (4)$$

$$\Rightarrow \frac{\Delta Q_\parallel}{n_{es}\sqrt{f_z/Z_{eff}}} \approx \sqrt{2\bar\kappa_o T_{es}^2 \int_0^{T_{es}} L_Z(T_e) T_e^{0.5} dT_e} \ \text{with} \ \kappa_o = \bar\kappa_o \frac{T_e^{2.5} Z_{eff}}{\ln\Lambda}$$

Table 3 Divertor radiation efficiencies for DIII-D and ITER.

| Element | Be | C | Ne | Ar | Kr |
|---|---|---|---|---|---|
| $f_Z$(%) for Zeff ~ 1.6 with $f_{He}$ = 0.1, $f_{Be}$ ~ 0.01 | 0.01 | 0.01 | .0033 | .001 | .00025 |
| $Q_\parallel$ITER/$\sqrt{(f_Z(\%)/Z_{eff})}$ at 250 eV (GW/m$^2$) | 0.08 | 0.2 | 0.7 | 2 | 4 |
| $Q_\parallel$ITER (coronal equilibrium) for $n_e \approx 10^{20}$ m$^{-3}$ | 0.065 | 0.16 | 0.33 | 0.52 | 0.52 |

These radiation levels can be enhanced by charge exchange recombination and rapid recycling of the impurities[6-8] and the radiation losses calculated as before (Figure 6). To radiate the required 100 MW in the divertor plasma takes relatively modest levels of recycling and neutral densities (Table 5).

    The power balance in Table 1 thus appears to be achievable for operation with 300 MW of ignited operation. Driven operation with 400 MW of total heating will require larger radiation levels from the divertor.

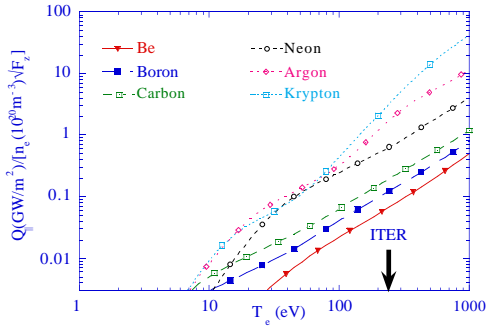 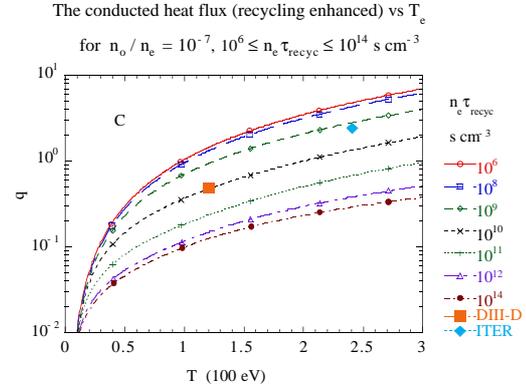

Figure 5. Radiation integral for parallel transport for a number of materials.

Figure 6. Recycling enhanced radiation integral for parallel transport for carbon.

Table 4  B2 calculations of the radiated power in the ITER divertor for two average Neon concentrations and upstream densities

| $n_e^{sep}$ ($10^{19}$ m$^{-3}$) | $f_{Ne}$ (%) | $P_{rad}$ (MW) |
|---|---|---|
| 4 | 0.5 | 70 |
| 4 | 0.25 | 40 |
| 6 | 0.5 | 100 |
| 6 | 0.25 | 60 |

Table 5 Enhanced Be, C, Ne, and Ar divertor radiation efficiencies for DIII-D and ITER.

| Element | Be | C | Ne | Ar |
|---|---|---|---|---|
| $f_Z$(%) for $Z_{eff} \sim 1.6$ with $f_{He} = 0.1$, $f_{Be} \sim 0.01$ | 1 | 1 | 0.33 | 0.1 |
| $Q_\|^{ITER}/\sqrt{(f_Z(\%)/Z_{eff})}$ at 250 eV required to radiate 100 MW | 0.8 | 0.8 | 1.4 | 2.5 |
| $n_o/n_e$ (required to radiate 100 MW) | $6 \times 10^{-3}$ | $10^{-3}$ | $3 \times 10^{-4}$ | $10^{-3}$ |
| $n_e\tau_{recycle}$ (s cm$^{-3}$)(required to radiate 100 MW) | $2 \times 10^{10}$ | $8 \times 10^{10}$ | $2 \times 10^{11}$ | $10^{10}$ |